\documentclass[12pt,preprint]{emulateapj}

\shorttitle{Implications for Short GRBs}
\shortauthors{Shao et al.}

\begin{document}

\title{Implications of Understanding Short Gamma-Ray Bursts Detected by {\it Swift}}

\author{\sc Lang Shao\altaffilmark{1,2,3}, Zi-Gao Dai\altaffilmark{4}, Yi-Zhong Fan\altaffilmark{1,2}, Fu-Wen Zhang\altaffilmark{1,2,5}, Zhi-Ping Jin\altaffilmark{1,2}, and Da-Ming Wei\altaffilmark{1,2}}
\altaffiltext{1}{Purple Mountain Observatory, Chinese Academy of Sciences, Nanjing 210008, China;lang@pmo.ac.cn}
\altaffiltext{2}{Key Laboratory of Dark Matter and Space Astronomy, Chinese Academy of Sciences, Nanjing 210008, China}
\altaffiltext{3}{Department of Physics, Hebei Normal University, Shijiazhuang 050016, China}
\altaffiltext{4}{Department of Astronomy, Nanjing University, Nanjing 210093, China}
\altaffiltext{5}{College of Science, Guilin University of Technology, Guilin 541004, China}

\begin{abstract}

In an effort to understand the puzzle of classifying gamma-ray bursts (GRBs), we perform a systematic study of {\it Swift} GRBs and investigate several short GRB issues. Though short GRBs have a short ($\lesssim2$~s) prompt duration as monitored by the Burst Alert Telescope, the composite light curves including both the prompt and afterglow emission suggest that most of the short GRBs have a similar radiative feature to long GRBs. Further, some well-studied short GRBs might also have an intrinsically long prompt duration, which renders them as a type of short GRB imposters. Genuine short GRBs detected by {\it Swift} might be rare that discriminating the observed short GRBs is, not surprisingly, troublesome. In particular, the observational biases in the host identification and redshift measurement of GRBs should be taken with great caution. The redshift distribution which has been found to be different for long and short GRBs might have been strongly affected by the measurement methods. We find that the redshifts measured from the presumed host galaxies of long and short GRBs appear to have a similar distribution.

\end{abstract}


\section{INTRODUCTION} \label{sec:intro}

Gamma-ray bursts (GRBs) are some of the most fascinating phenomena in the universe with profound physical implications hidden in its diverse prompt and afterglow emission. Phenomenologically, there are two types of GRBs (long versus short) based on the bimodal distribution of their prompt duration \citep{kouveliotou93}. The occasional discovery of several nearby long GRBs associated with Type Ic supernovae (SNe) favors the speculation that most long GRBs are accompanied by massive stellar explosions \citep[see][for a review]{woosley06}. Short GRBs with a typical duration of about tenths of a second are generally considered unlikely to result from the death of massive stars which would have a typical timescale (the free-fall time) of tens of seconds; instead, they are proposed to result from compact mergers \citep[see][for a review]{nakar07}. The compact progenitor models generally predict a lower star-forming environment, and have been supported by the fortuitous localization of a few nearby short GRBs \citep{fox05,gehrels05,berger05a,barthelmy05}.

This dichotomous classification scheme based on prompt duration was soon found controversial after the discovery of a nearby event, GRB060614, which has a long duration but shows clear evidence of no accompanying SN \citep{fynbo06,dellavalle06} and has an initial hard pulse that exhibits the same properties as short GRBs \citep{gehrels06,zhang07}. Ever since then, the question, whether these two types of GRBs have different origins and how to categorize individual GRB have remained open \citep[][and references therein]{zhang09}.

In this paper, we perform a systematic study of both long and short GRBs detected by the {\it Swift} satellite, and investigate several issues that would help refresh our understanding of short GRBs. This paper is structured as follows. In Section 2, we make a comparison between the composite light curves of 14 well-monitored short GRBs and 137 long GRBs in their rest frame. In Section 3, we revisit the relation between the prompt duration and the spectral hardness ratio for 500 {\it Swift} GRBs that used to play an important role in distinguishing between long and short GRBs. In Section 4, we revisit the redshift distribution that has been found to be different between long and short GRBs. In Section 5, we give a summary of our conclusions.

\section{COMPOSITE BAT-XRT LIGHT CURVES} \label{sec:lightcurve}

Some of the difficulties in categorizing a {\it Swift} GRB come from the fact that (1) both the prompt and afterglow emission are very diverse in their observational properties and there is {\it no} unambiguous definition for the dividing line between them so that the prompt duration cannot be accurately determined without contamination from the afterglow and (2) they are usually studied in separate bandpasses of two different detectors, i.e., the Burst Alert Telescope (BAT) for the prompt duration and the X-Ray Telescope (XRT) for the afterglow, with the prompt duration appearing to be strongly dependent on the bandpass and sensitivity of the detector.

These difficulties might be overcome with the BAT-XRT composite light curves produced by extrapolating BAT and XRT data into a single bandpass \citep{obrien06,sakamoto07,evans10}. The BAT and XRT light curves can usually join smoothly, and the distinction between the prompt and afterglow emission is usually manifest in a prompt plateau followed by a monotonically decaying afterglow \citep{obrien06,shao10}. Even though the prompt and afterglow emission are undistinguishable for some bursts with late X-ray flares, the duration of the prompt plateau is generally consistent with the $T_{90}$ evaluated by BAT for most long GRBs \citep{shao10}. For short GRBs, we expect that the duration of their prompt plateau be short (say $\lesssim2$~s) as well. The overall composite light curve is also an important factor in understanding the physical origin of short GRBs.

\begin{figure}
\begin{center}
\includegraphics[width=0.6\textwidth]{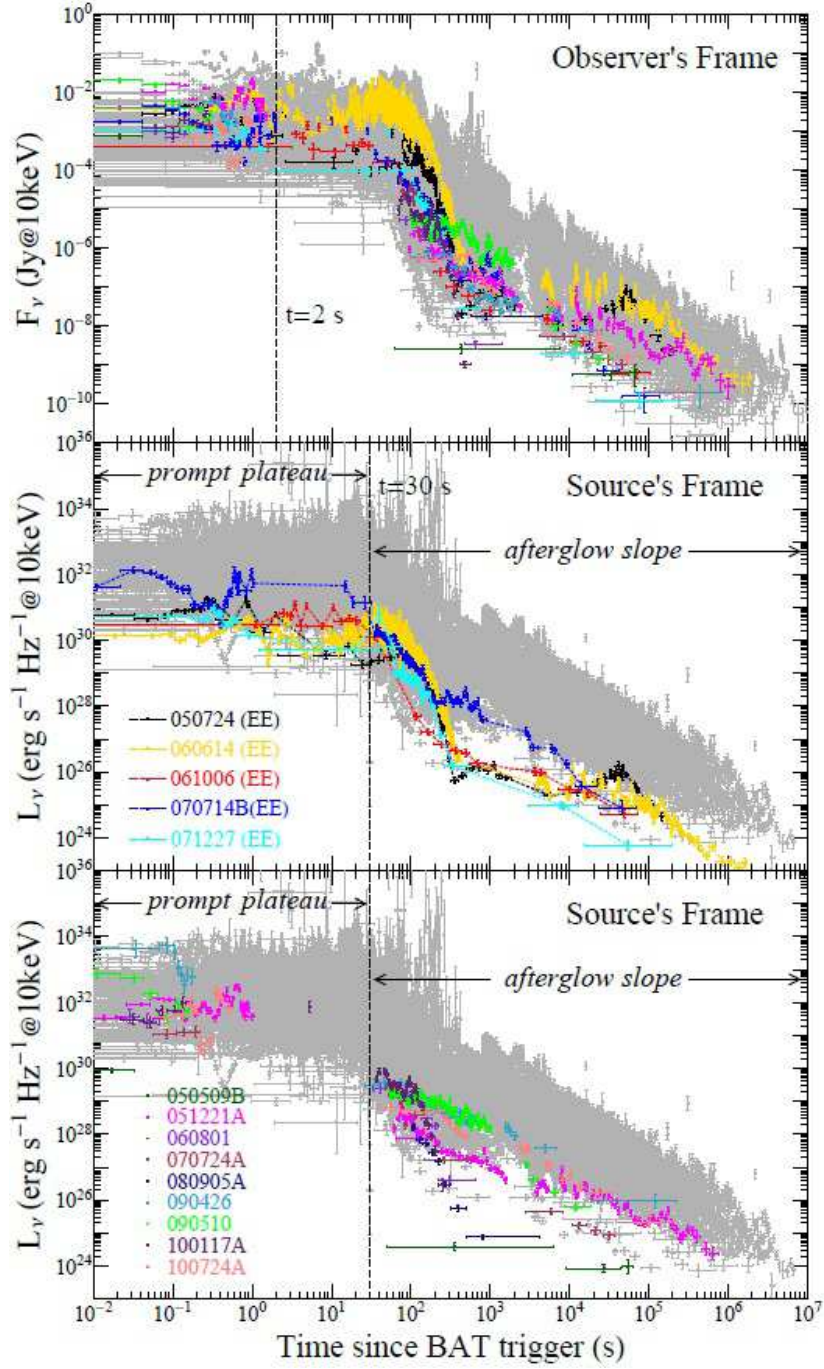}
\end{center}
\caption{X-ray light curves of 137 long (gray) and 14 short (colored) GRBs in the observer frame (top) and the rest frame (middle and bottom). Five events known as short GRBs with extended emission (EE) are shown in the middle panel and other nine short GRBs are shown in the bottom panel. }
\label{fig:lc}
\end{figure}

We make extensive use of the automated {\it Swift} Burst Analyser provided by the UK {\it Swift} Science Data Centre \citep{evans10}. As listed in Table~\ref{tab:sample}, we acquire a sample of 14 GRBs well monitored by {\it Swift} (up to GRB 100724A), which have both a composite light curve and a measured redshift and have been well studied in the literature as short GRBs. In particular, five of them have an emission component in BAT known as extended emission \citep[EE; e.g.,][]{norris06}. The composite light curves of these 14 short GRBs and 137 long GRBs are compared both in the observer and source frames as shown in Figure~\ref{fig:lc}. For the last BAT data point and the first XRT data point of these short GRBs, the time since BAT trigger in the observer frame is also given in Table~\ref{tab:sample}.

The 137 long GRBs are taken from the sample of a total of 150 GRBs in \citet{shao10}. Among the remaining 13 of those 150 GRBs, 10 are included in this work as short GRBs and 3 (GRB~060218, 090417B, and 100316D) are dropped as being peculiar: GRB~060218 and 100316D are extremely underluminous with associated SNe \citep[e.g.,][]{soderberg06a,starling11,fan11}; GRB~090417B is unique because it has been discovered to have significant spectral softening \citep{holland10} which reveals the potential existence of an additional radiative component \citep{shao07,shao08}. For details of how these light curves were produced, see \cite{evans10} and \cite{shao10}. The figures in this work are produced with the LevelScheme package for Mathematica \citep{caprio05}.

{\it As a group}, shown in the observer's frame (top panel of Figure~\ref{fig:lc}), the overall light curves of these short GRBs clearly fall into the region occupied by long GRBs, suggesting a similar radiative feature as long GRBs. Even though only five of them have detectable prompt emission after 2 s, the existence of an intrinsically similar prompt plateau with a duration of tens of seconds, similar to long GRBs, might not be ruled out. After being corrected for the cosmological effects based on the reported redshifts listed in Table~\ref{tab:sample}, the light curves in the source frame also reveals the same similarities (middle and bottom panels of Figure~\ref{fig:lc}). However, in the source frame, these short GRBs appear to be less luminous than most long GRBs\footnote{What causes the contrast between the observer frame and the source frame? Technically, the difference is due to a relatively lower redshift distribution for short GRBs. As we will discuss in Section 4, the redshift distribution might have been strongly affected by observational biases.}.

{\it As individuals}, some short GRBs also exhibit the transition of a ``prompt plateau'' to an ``afterglow slope'' at approximately tens of seconds, similar to that proposed by \citet{shao10} for most long GRBs. As shown in the middle panel of Figure~\ref{fig:lc}, these cases are usually considered as short GRBs with EE. The uncertainties in extrapolating the BAT data into the XRT band usually arise from the discontinuity of photon index between BAT and XRT and might play an important role in shaping the composite light curves \citep{sakamoto07}. However, this transition usually does not change dramatically within different energy ranges, the discontinuity of the photon index is only present in about $20\%$ of cases, and the flux density is less affected by this discontinuity since it lies close to both bandpasses \citep{evans10}. In particular, the duration of the plateau phase is in general consistent with the $T_{90}$ evaluated by BAT \citep{shao10}. Note that we have disregarded unreliable data indicated with ``!\#'', which are caused by the uncertainties in extrapolation \citep{evans10}.

With the transition from a ``prompt plateau'' to an ``afterglow slope'' clearly showing in some GRBs, the real prompt duration could be more accurately inferred with less biases caused by the observational limitations of the prompt detectors. For this reason, most events like GRB 060614, shown in the middle panel of Figure~\ref{fig:lc}, probably have an intrinsically long (say, longer than 2 s) prompt duration. Based on the spectral properties, these short GRBs with EE are also very similar to long GRBs, as we will discuss in the next section \citep[see also][]{sakamoto11}. As suggested by the last BAT data point in the bottom panel of Figure~\ref{fig:lc} (see also Table~\ref{tab:sample}), GRB 100117A might also have an EE component, which reveals its identity as a short GRB imposter, with most of its prompt emission probably hidden slightly below the background level.

On the other hand, if some short GRBs indeed have a short prompt duration \citep[e.g.,][]{norris10}, this transition should be expected as well. Unfortunately, as shown in the bottom panel of Figure~\ref{fig:lc}, this case has never been clearly established for the short GRBs detected by {\it Swift} due to the lack of sufficient observational data. The most ideal one would be GRB~090426 \citep{thoene09,antonelli09,levesque10,xin11}, but it might still be risky to conclude based on current data. On the contrary, recent works suggest that this burst has some properties very similar to that of long GRBs \citep{thoene11,nicuesa11}. Further investigation of the prompt-to-afterglow transition and the confirmation of an intrinsic short GRB based on this transition will be worth looking forward to in future broadband GRB missions, e.g., the Space multi-band Variable Object Monitor \citep{gotz09}. On a side note, it is unclear why the transition for most GRBs at tens of seconds seems to match nicely the commencement of XRT data (see the bottom panel of Figure~\ref{fig:lc}).

\section{PROMPT DURATION AND HARDNESS RATIO} \label{sec:anti}

\subsection{DISTRIBUTION OF DURATION} \label{sec:dura}

\begin{figure}
\begin{center}
\includegraphics[width=0.6\textwidth]{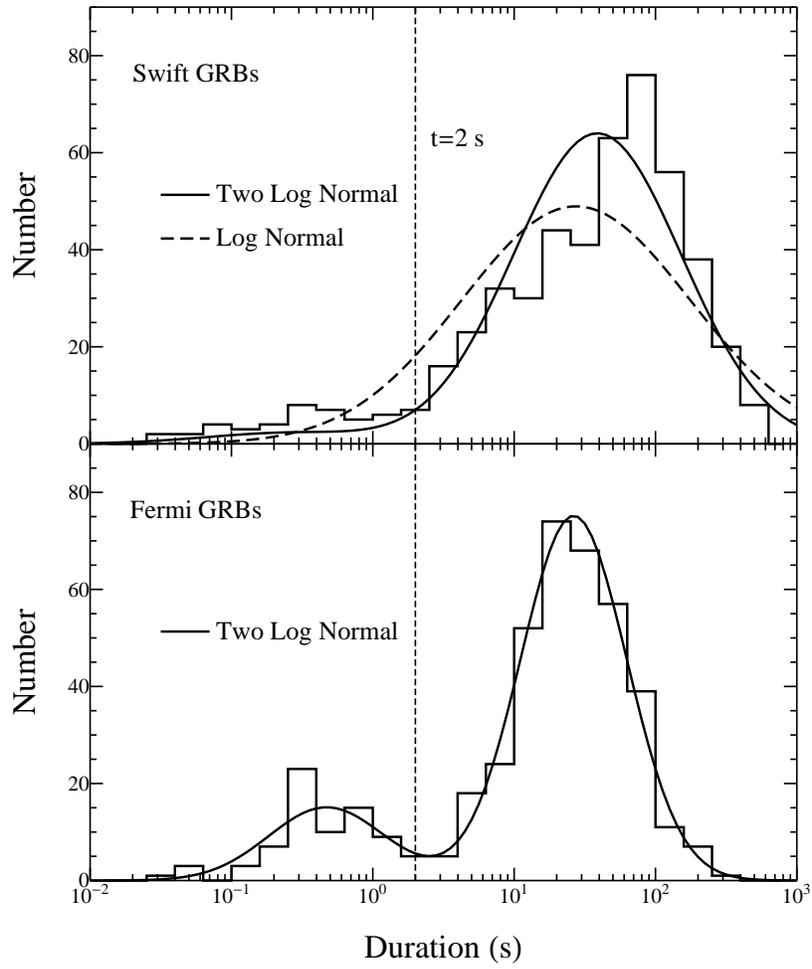}
\end{center}
\caption{Top panel: distributions of the observed prompt durations of 495 GRBs detected by {\it Swift}. Bottom panel: distribution of the observed prompt durations of 432 GRBs detected by {\it Fermi} up to GRB 100330856 \citep[data from][]{nava11}.}
\label{fig:t90}
\end{figure}

As is well known, the primary motivation for discriminating long GRBs from short GRBs at the separation line of $\sim2$~s was triggered by the bimodal distribution of the prompt durations of BATSE GRBs \citep{kouveliotou93}. This bimodal distribution had been previously proposed for KONUS GRBs \citep{mazets81} and has been recently confirmed for {\it Fermi} GRBs \citep{nava11}. As shown in the bottom panel of Figure~\ref{fig:t90}, the distribution of 432 {\it Fermi} GRBs up to GRB 100330856 \citep{nava11} can be well fitted by two lognormal distributions which have a mean $\mu_1=-0.33$ ($\sim0.47$~s; with a standard deviation $\sigma_1=0.41$ and a weight $w_1=0.18$) and a mean $\mu_2=1.42$ ($\sim26.37$~s; with a standard deviation $\sigma_2=0.38$ and a weight $w_2=0.82$). The probability ({\it p}-value) for this fit is $99.16\%$ and a third lognormal distribution might not be needed \citep{horvath98}. The expected percentage of short GRBs should be $w_1=17.98\%$ and the number of GRBs that have a duration less than 2~s in this sample is 73 (corresponding to a rate of $16.90\%$). These statistical results are consistent with those of BATSE GRBs \citep{kouveliotou93,meegan96}.

However, the bimodal distribution is not manifest for {\it BeppoSAX} GRBs \citep{frontera09} and {\it Swift} GRBs \citep{sakamoto08,sakamoto11}.  As shown in the top panel of Figure~\ref{fig:t90}, the prompt durations of 495 {\it Swift} GRBs up to GRB100904A could not be well fitted by either one lognormal or two lognormal distributions. The best fit with a single lognormal has a mean $\mu=1.44$ ($\sim27.34$~s; with a standard deviation $\sigma=0.81$), but the probability for this fit is only $0.002\%$. Meanwhile, the probability for two lognormals, which have a mean $\mu_1=-0.57$ ($\sim0.27$~s; with a standard deviation $\sigma_1=0.60$ and a weight $w_1=0.04$) and a mean $\mu_2=1.59$ ($\sim38.84$~s; with a standard deviation $\sigma_2=0.60$ and a weight $w_2=0.96$), respectively, is better but still only $1.77\%$. According to this two-lognormal distribution, the percentage of short GRBs is only $w_1=3.51\%$. In fact, the number of GRBs with a duration less than 2~s among the 495 {\it Swift} GRBs is 43, among which at least 8 should belong to the tail of the long GRBs distribution. Therefore, the real percentage of short GRBs with a duration less than 2~s in the 495 {\it Swift} GRBs would be less than $\sim7\%$ \citep{sakamoto11}.

Technically, the discrepancy on the distribution of the prompt duration might be due to the lower efficiency of triggering on a short GRB due to the stringent triggering criteria of these monitors \citep{sakamoto08,frontera09}. In fact, as we have shown in Section~\ref{sec:lightcurve}, even for those short GRBs that have already been detected by {\it Swift}, it is still very likely that some, though not all, of them have an intrinsically long prompt duration \citep[e.g.,][]{norris10}. Even if the debatable bimodality of the prompt duration has revealed the ``tip of the iceberg'' as to two types of GRBs with different characteristic timescales \citep{kouveliotou93}, the lack of short GRBs and the existence of (probably) considerable short GRB imposters would have made {\it Swift} GRBs a severely biased sample in favor of only long GRBs. Therefore, it is not surprising that one would face a dilemma when classifying {\it Swift} GRBs.

\subsection{CORRELATION BETWEEN DURATION AND HARDNESS} \label{sec:corr}

\begin{figure}
\begin{center}
\includegraphics[width=0.8\textwidth]{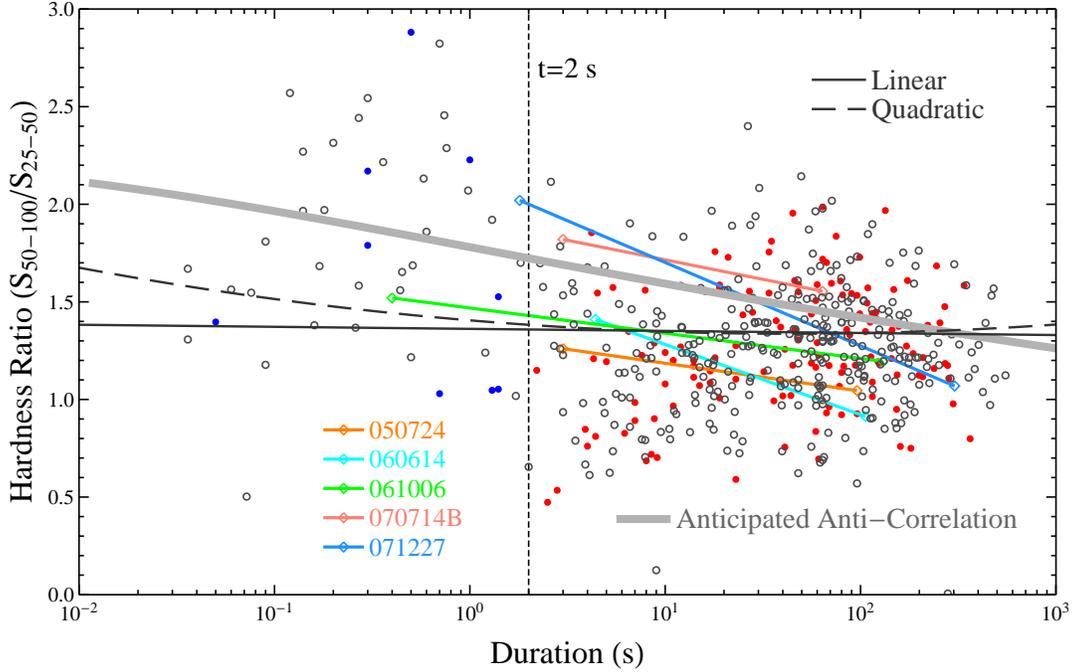}
\end{center}
\caption{Hardness ratio and observed duration of 500 GRBs detected by {\it Swift} up to GRB 100904A. Each of the five short GRBs with extended emission (EE) is treated as two GRBs connected with a solid line: one with the initial hard spike only and one with the soft EE included \citep{zhang09}. The latter fall wells into the region occupied by long GRBs. The other nine short GRBs are marked with blue dots. The 137 long GRBs shown in Figure~\ref{fig:lc} are marked with red dots. The rest are marked with open gray circles. The solid and dashed lines represent the best fits by linear and quadratic functions, respectively. The thick gray line represents the anticipated anti-correlation given by Equation~(\ref{eq:hdratio}). All the error bars are hidden for better visibility.}
\label{fig:hdrt90}
\end{figure}

The anti-correlation between the prompt duration and the spectral hardness ratio was considered as the missing link that showed evidence of the physical difference between the long and short GRBs detected by BATSE \citep{kouveliotou93}. The {\it Swift} detection of soft EE in several short GRBs (very likely imposters) has complicated this issue but might also help refresh our understanding of this issue. As shown in Figure~\ref{fig:hdrt90}, all five short GRBs fall into the region of long GRBs if their soft EE is included as the prompt emission \citep[see also][]{sakamoto11}. For these cases, we would reproduce an anti-correlation between the hardness ratio and the prompt duration, even assuming that all GRBs are intrinsically similar.

In fact, it has been proposed that both the temporal and spectral properties of some short GRBs are probably similar to those of the first 1-2~s of long GRBs detected by BATSE \citep{nakar02,ghirlanda09}. In addition, the prompt spectral evolution of some short GRBs has also been proposed to be consistent with that of long GRBs detected by {\it Fermi} \citep{ghirlanda11}. As we have proposed in our previous work for {\it Swift} GRBs, the prompt emission of long GRBs usually exhibit weak or negligible softening in the first few seconds and then undergo severe softening later \citep{shao10}. Therefore, if the prompt detector only caught the first 2~s of a long GRB, it would be probably taken as a short and hard GRB, and its spectral lag would also be negligible compared to that of the long GRB \citep{norris06}.

To quantitatively describe this anti-correlation, for simplicity, we shall evaluate the average hardness ratio over different bursts by
\begin{equation}
<{\rm HR}_{32}(T)>=<{S_3(T)\over S_2(T)}>=<{\int_0^T\int_{\rm 50 keV}^{\rm 100 keV}F_E{\rm d}E{\rm d}t \over \int_0^T\int_{\rm 25 keV}^{\rm 50 keV}F_E{\rm d}E{\rm d}t}>,\label{eq:hd0}
\end{equation}
where ${\rm HR}_{32}$ is the ratio of fluences in the third channel ($S_3$) and second channel ($S_2$) of {\it Swift}/BAT \citep{gehrels04}, $F_E$ is the flux density, $T$ is the observed GRB duration and $t$ is the time since BAT trigger. In general, the spectra of most GRBs can be fitted with a power-law model in the BAT bandpass \citep{sakamoto08,sakamoto09,evans10}, so we have
\begin{equation}
<F_E>\propto \left({E\over 10\,{\rm keV}}\right)^{1-<\Gamma(t)>},
\end{equation}
and the evolution of the average photon index may have the form
\begin{equation}
<\Gamma(t)>=\Gamma_0+\delta\cdot\log\left({t \over 10^{-2}\,{\rm s}}\right),
\end{equation}
where the averages are evaluated over different bursts and we have $\Gamma_0\sim0.92$ and $\delta\sim0.17$, obtained by \citet{shao10}. Therefore, we can have an anticipated anti-correlation between the hardness ratio and prompt duration given by (thick gray line in Figure~\ref{fig:hdrt90})
\begin{equation}
<{\rm HR}_{32}(T)>= {\int_{10^{-2}}^{T+10^{-2}}\int_{\rm 50 keV}^{\rm 100 keV}(E/10\,{\rm keV})^{1-<\Gamma(t)>}{\rm d}E{\rm d}t \over \int_{10^{-2}}^{T+10^{-2}}\int_{\rm 25 keV}^{\rm 50 keV}(E/10\,{\rm keV})^{1-<\Gamma(t)>}{\rm d}E{\rm d}t},\label{eq:hdratio}
\end{equation}
where the starting point for the integral over time can be set at $10^{-2}\,{\rm s}$, keeping the evaluation valid and implicitly assuming that the initial spike (with a typical timescale of $10^{-2}\,{\rm s}$) could always be detected.

We have shown above that the anti-correlation that was first proposed by \citet{kouveliotou93} may be easily reproduced and has been exemplified by these short GRBs with EE as can be seen in Figure~\ref{fig:hdrt90}. However, this anti-correlation might not be universal. Pearson's correlation coefficient between the hardness ratio and the prompt duration for the 495 {\it Swift} GRBs in our work is only $-0.23$. In particular, there is no anti-correlation or there is even very weak positive correlation when long and short GRBs are evaluated separately, since the correlation coefficients are 0.11 and 0.08, respectively. This negligible correlation is consistent with that is found for {\it BeppoSAX} GRBs \citep{frontera09}.

To explore the potential existence of a more realistic structure in the hardness ratio and prompt duration diagram \citep[e.g.,][]{horvath10}, as shown in Figure~\ref{fig:hdrt90}, we carry out series of model fits with linear, quadratic, cubic, quartic, quintic functions, etc. For instance, for a linear model $f(x)=ax+b$, the best-fit parameters are $a=-0.01\pm0.04$ and $b=1.36\pm0.08$. For a quadratic model $f(x)=ax^2+bx+c$, the best-fit parameters are $a=0.02\pm0.04$, $b=-0.08\pm0.14$ and $c=1.41\pm0.12$, and so on. As the number of free parameters is increased, the diverse feature in short GRBs and the coherent feature in long GRBs are revealed. No sub-groups could be confirmed due to the severe overlap \citep{sakamoto11}.

\begin{figure}
\begin{center}
\includegraphics[width=0.8\textwidth]{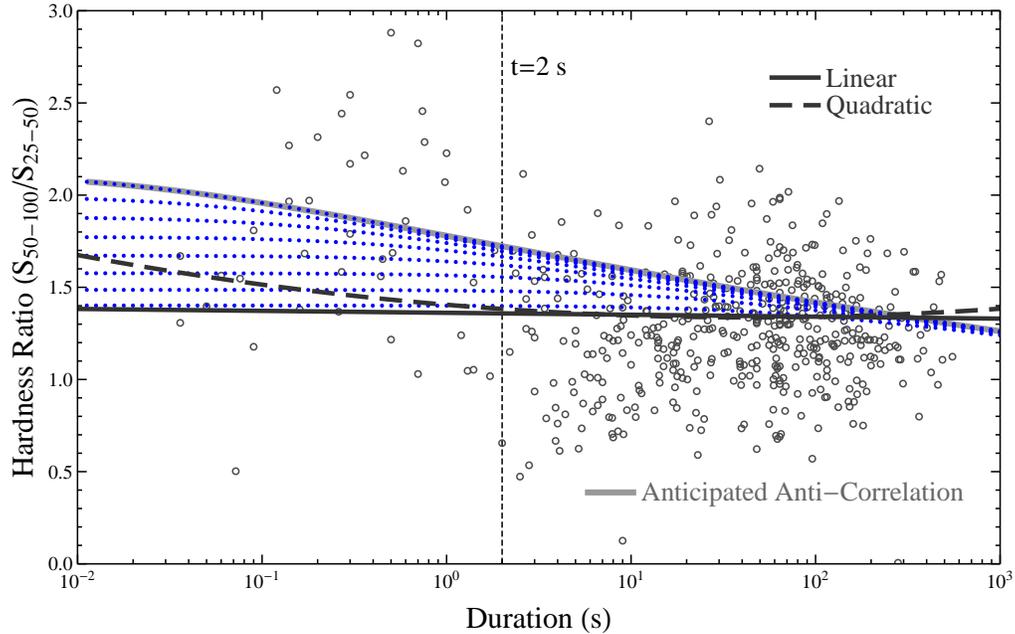}
\end{center}
\caption{Same as Figure~\ref{fig:hdrt90}. The blue dots represent the evaluated hardness ratios as given by Equation~(\ref{eq:hdratio2}) as the free parameter $t_0$ varies from $10^{-2}$ to $30$~s (from top to bottom with a step of $\times 10^{0.5}$) for each given duration $T$. All the error bars are hidden for better visibility.}
\label{fig:hdrt902}
\end{figure}

The weak or negligible anti-correlation might be accounted for if a more realistic scenario is considered. Again, for simplicity, we propose a modification to Equation~(\ref{eq:hdratio}) with the new form
\begin{equation}
<{\rm HR}_{32}(T)>= {\int_{t_0}^{t_0+T}\int_{\rm 50 keV}^{\rm 100 keV}(E/10\,{\rm keV})^{1-<\Gamma(t)>}{\rm d}E{\rm d}t \over \int_{t_0}^{t_0+T}\int_{\rm 25 keV}^{\rm 50 keV}(E/10\,{\rm keV})^{1-<\Gamma(t)>}{\rm d}E{\rm d}t},\label{eq:hdratio2}
\end{equation}
where the starting point for the integral over time $t_0$ is an additional free parameter which represents the time of the BAT trigger since the real onset of the GRB (also assuming that all GRBs are intrinsically similar). This is to reflect the cases where the spike triggering BAT is not the first spike of the GRB, and the preceding spikes have been hidden slightly below the background level. Given that the prompt spikes are diverse and the triggering criteria are complicated, this triggering time $t_0$ might not be well constrained. As shown in Figure~\ref{fig:hdrt902}, given a certain $T$, the hardness ratio gets smaller as $t_0$ gets larger. As $t_0$ gets as large as $\sim 30$~s, the hardness ratio becomes almost uncorrelated with $T$. In particular, long GRBs are found to be much more immune to this effect, which is qualitatively consistent with the fact that they are more coherent (see also the distribution of the hardness ratio discussed later in this section). This implies that some short GRBs with a soft prompt spectrum similar to the long GRBs could be an event only containing a later (but bright) spike of a long GRB. As suggested by the two lower panels of Figure~\ref{fig:lc}, some long GRBs tend to have a peak luminosity at tens of seconds.

\begin{figure}
\begin{center}
\includegraphics[width=0.6\textwidth]{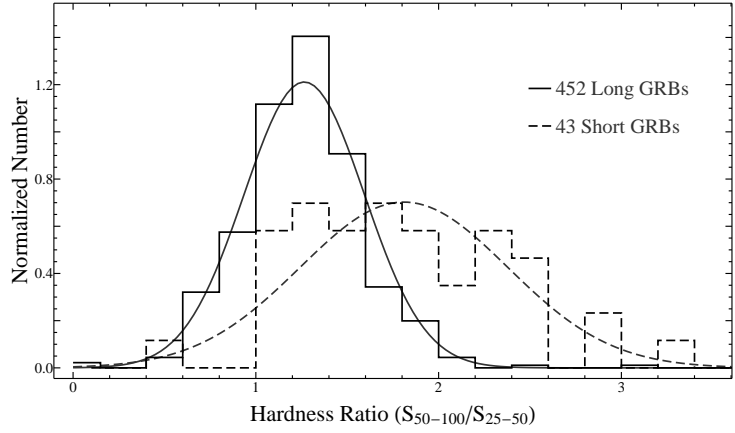}
\end{center}
\caption{Distribution of the hardness ratio of 452 long GRBs and 43 short GRBs detected by {\it Swift} up to GRB 100904A. The solid and dashed lines show the best fits with a normal distribution.}
\label{fig:hdr}
\end{figure}

For completeness, we also analyze the distribution of the hardness ratio of long and short GRBs separately. As shown in Figure~\ref{fig:hdr}, the distribution of the hardness ratio of long GRBs can be well fitted by a relatively tight normal distribution, with a mean $\mu=1.26$ and a standard deviation $\sigma=0.33$. However, the distribution of the hardness ratio of short GRBs has a large dispersion, and the Gaussian fitting yields a mean $\mu=1.81$ and a standard deviation $\sigma=0.57$. This diverse feature in the hardness ratio of short GRBs also reveals that a considerable fraction of short GRBs detected by {\it Swift} is not intrinsic \citep{norris10,norris11}.

\section{REDSHIFT DISTRIBUTION} \label{sec:redshift}

\begin{figure}
\begin{center}
\includegraphics[width=0.6\textwidth]{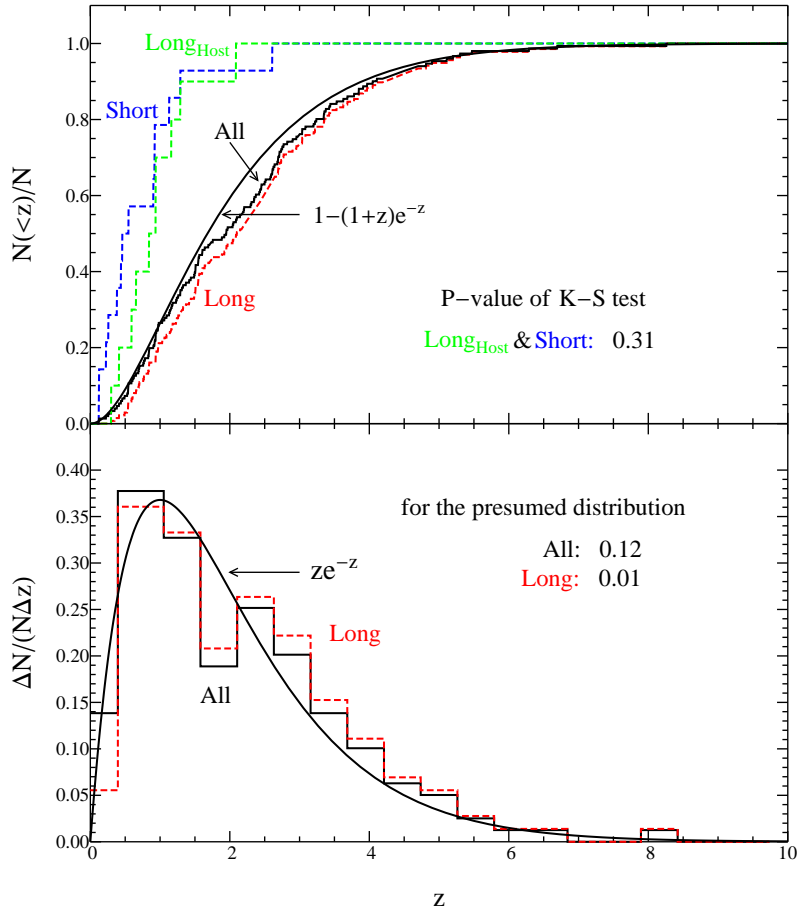}
\end{center}
\caption{Redshift distribution of 137 long and 14 short GRBs. Top panel: the cumulative distribution function (CDF) for short (blue dashed steps) and long (red dashed steps) GRBs and their sum (black steps). The cumulative distribution of the redshifts of the 10 long GRBs measured from their presumed host galaxies is also shown in green dashed steps. The solid curve is the CDF of the presumed Erlang distribution with shape parameter $k=2$ and rate parameter $\lambda=1$.  Bottom panel: the probability density function (PDF) for the 151 GRBs (black histogram) and the 137 long GRBs (red dashed histogram). The solid curve is the PDF of the same Erlang distribution. }
\label{fig:redshift}
\end{figure}

The host galaxies of short GRBs detected by {\it Swift} are found to be different from those of long GRBs \citep{berger09}, and the localization of a few short GRBs has indicated that they tend to be in the outer region of the host galaxy \citep{fox05,gehrels05,berger05a,barthelmy05,berger10}, which is in stark contrast to the cases of long GRBs. As shown in the top panel of Figure~\ref{fig:redshift}, the redshift distribution of both types of GRBs in our sample (blue dashed steps and red dashed steps) is also significantly different \citep{berger09}. However, these evidences should be taken with great caution because of the uncertainty of host identification \citep{cobb06,cobb08}. Even though the possibility of the coincidental superposition of a GRB and a foreground galaxy has been found to be quite small \citep[$\sim 1\%$;][]{cobb08}, this rareness might have always been taken as a wise argument for confirming host identification even when there may be a significant offset from some short GRBs. As more and more GRBs are detected in the future, the uncertainties would be finally exposed.

Here we focus on the redshift distribution, which might have been affected by the observational biases. First, as is well known, the redshift of a GRB could be measured by the spectral lines either from its afterglow or from its host galaxy, with the detection of the host galaxy strongly favored at a lower redshift. In fact, for the measured redshifts of 14 short GRBs in our sample, 12 of them are inferred from the spectral lines of the presumed host galaxies. The two exceptions are GRB 090426 \citep[$z=2.609$;][]{levesque10} and 100724A \citep[$z=1.288$;][]{thoene10}, the redshifts of which are inferred from the spectral lines of their afterglows. For the measured redshifts of 137 long GRBs in our sample, only 10 of them are inferred from the presumed host galaxies as listed in Table~\ref{tab:sample2}, which also tend to have a lower redshift. Interestingly, the Kolmogorov-Smirnov (K-S) test suggests that the possibility that the redshifts of these short GRBs and these long GRBs measured from presumed host galaxies (the green dashed steps in the top panel of Figure~\ref{fig:redshift}) have been drawn from the same distribution could be as high as $\sim31\%$. Therefore, the discrepancy of the redshift distribution between long and short GRBs should be taken with caution.

Second, there are many obstacles to obtaining a GRB redshift, making it a rare event. Regardless, the present redshift distribution has been considered an important indicator of the GRB rate or star-forming rate of the host galaxies \citep{le07,berger09} and has been proposed as a good tracer for star formation in the universe up to a high redshift \citep{jakobsson06}. However, we propose here that these results should be taken with caution. As shown in Figure~\ref{fig:redshift}, the joint redshift distribution of the 137 long GRBs and 14 short GRBs appears to be asymptotic to a parameter-free mathematical distribution,
\begin{equation}
F(x)=1-(1+x)e^{-x}, \label{eq:cdf}
\end{equation}
for the cumulative distribution function (CDF) and to
\begin{equation}
f(x)=F'(x)=xe^{-x}. \label{eq:pdf}
\end{equation}
for the probability density distribution (PDF). The K-S test for the null hypothesis that the redshifts of the 137 long GRBs follow this distribution might be rejectable, with a probability value ({\it p}-value) of $0.01$\footnote{Conventionally, the condition to accept the null hypothesis is if the {\it p}-value$>0.05$.}. However, the null hypothesis that the redshifts of the 151 GRBs (with 137 long and 14 short GRBs together) follow this distribution is more acceptable with a {\it p}-value of $0.12$. This distribution without any parameter suggests a Poissonian (or rare and random) nature of the GRB redshift. Indeed, it might be closely related with the so-called Erlang distribution,
 \begin{equation}
F(x;k,\lambda)=1-\sum_{n=0}^{k-1}e^{-\lambda x}(\lambda x)^n/n!, \label{eq:erlang}
\end{equation}
with shape parameter $k=2$ and rate parameter $\lambda=1$, known as the probability distribution of the waiting time until the second ``arrival'' in a one-dimensional Poisson process with a given rate \citep[e.g.,][]{billingsley86}. A further investigation into the nature of this distribution is in need.

It is worth to mention that, as we have shown in Figure~\ref{fig:lc}, the short GRBs detected by {\it Swift} appear to have an overall lower luminosity in the source frame, even though most of them appear to be comparable with long GRBs in the observer frame. The major difference is due to the relatively lower redshifts determined for short GRBs. For a short GRB like GRB050509B that appears to be remarkably underluminous \citep[e.g.,][]{shao10} and a little far away from the center of the presumed host galaxy at redshift $z=0.225$ \citep{gehrels05}, the presumed host galaxy and the corresponding redshift might be questionable.

The joint redshift distribution also suggests that the current techniques for determining GRB redshift are biased in favor of lower ones. For instance, according to Equation~(\ref{eq:cdf}), the expected number of redshifts between $z=6-10$ for a sample of 151 measured redshifts should only be $\sim2.54$ (i.e., $151\times[F(10)-F(6)]\simeq2.54$). Accordingly, we indeed have three bursts in our sample located between $z=6-10$ (GRB 050904: $z=6.29$; GRB 080913: $z=6.70$; GRB 090423: $z=8.26$). Of course, we expect this distribution to be much improved toward larger redshift with advanced techniques extensively involved in the future \citep[e.g.,][]{postigo10}.

\section{CONCLUSION} \label{sec:conclusion}

In this work, we performed a systematic study of {\it Swift} GRBs and address several issues that may help refresh our understanding of short GRBs and the current puzzle of classifying GRBs detected by {\it Swift}. Here is a summary of our conclusions.

1. Though some short GRBs have a short prompt duration as observed by BAT, the BAT-XRT composite light curves suggest that most of them may have an overall similar radiative feature to long GRBs, which has also been suggested by recent work on the short GRB detected by {\it Fermi} \citep{ghirlanda11}.

(2) As also suggested by the composite light curves, some well-studied short GRBs detected by {\it Swift} may also have an intrinsically long prompt duration, which renders them as a type of short GRB imposters, exemplified by some short GRBs with EE. We propose that genuine short GRBs detected by {\it Swift} might be rare.

(3) The observational biases in the host identification and redshift measurement should be taken with great caution. The redshift distribution which has been found to be different for long and short GRBs might have been strongly affected by the measurement methods. Short GRBs tend to have lower redshift, very similar to those of long GRBs measured by the same method, i.e., spectral analysis of the presumed host galaxies.

\acknowledgments

We are very grateful to the referee for insightful comments and suggestions. This work made use of data supplied by the UK {\it Swift} Science Data Centre at the University of Leicester. We are grateful to S.~D.~Barthelmy and T.~Sakamoto for the GCN data and J.~Greiner for online GRB table. L.S. acknowledges helpful discussion with M.~Caprio on the LevelScheme package for Mathematica. This work was supported by the National Natural Science Foundation of China (grants 10673034, 10621303, 10873009, 11033002 and 11073057) and the National Basic Research Program of China (Nos.~2007CB815404 and 2009CB824800). F.W.Z. acknowledges the support by Guangxi Natural Science Foundation
(2010GXNSFB013050) and the doctoral research foundation of Guilin University of Technology.

\clearpage

\begin{table}
\caption{Short GRBs with a Measured Redshift}
\begin{center}
\begin{tabular}{|cl|l|c|c|c|c|}
\hline
&GRB          & $T_{90}$$^b$ & $z$ & Last Data  & First Data  & Ref.$^d$ \\
&        & (s) &  & of BAT (s) & of XRT (s) & \\
\hline
&050509B      & 0.05    &  0.2248   & 0.02 & 429.4 &1,2  \\
&050724$^a$   & 3$^c$   &  0.257    & 103.3 & 79.5 &3--5  \\
&051221A      & 1.40    &  0.5465   & 1.5 & 96.0 &6,7  \\
&060614$^a$   & 4.4$^c$ &  0.125    & 198.9 & 97.5 &5,8,9  \\
&060801       & 0.50    &  1.1304   & 0.1 & 87.1 &10,11  \\
&061006$^a$   & 0.4$^c$ &  0.4377   & 51.0 & 168.3 &5,11,12 \\
&070714B$^a$  & 3$^c$   &  0.9224   & 49.7 & 68.6 &5,13  \\
&070724A      & 0.66    &  0.4571   & 0.3 & 74.9 &14  \\
&071227$^a$   & 1.8$^c$ &  0.381    & 80.2 & 86.4 &5,12,14  \\
&080905A      & 1.0     &  0.1218   & \nodata & 116.2 &15  \\
&090426       & 1.25    &  2.609    & 0.6 & 124.4 &16--19 \\
&090510       & 0.30    &  0.903    & 0.3 & 99.8 &20--22  \\
&100117A      & 0.29    &  0.92     & 10.0 & 70.7 &23  \\
&100724A      & 1.39    &  1.288    & 1.2 & 126.5 &24  \\
\hline
\end{tabular}
\end{center}
\begin{minipage}{18cm}
$^a$ Bursts that have known extended emission (EE). \\
$^b$ The duration that encompasses $90\%$ of the total GRB counts detected by {\it Swift} BAT. \\
$^c$ The duration that only encompasses the first short and hard pulse and omits EE. \\
$^d$ References.
(1)~\citealt{gehrels05}; (2)~\citealt{bloom06}; (3)~\citealt{barthelmy05}; (4)~\citealt{berger05a}; (5)~\citealt{zhang09}; (6)~\citealt{soderberg06b}; (7)~\citealt{burrows06}; (8)~\citealt{price06}; (9)~\citealt{zhang07}; (10)~\citealt{cucchiara06}; (11)~\citealt{berger07}; (12)~\citealt{davanzo09}; (13)~\citealt{graham09}; (14)~\citealt{berger09}; (15)~\citealt{rowlinson10}; (16)~\citealt{thoene09}; (17)~\citealt{antonelli09}; (18)~\citealt{levesque10}; (19)~\citealt{xin11}; (20)~\citealt{rau09}; (21)~\citealt{depasquale10}; (22)~\citealt{mcbreen10}; (23)~\citealt{fong11}; (24)~\citealt{thoene10}.
\end{minipage}
\label{tab:sample}
\end{table}


\begin{table}
\caption{Long GRBs with a Measured Redshift of the Presumed Host Galaxy}
\begin{center}
\begin{tabular}{|cl|l|c|c|}
\hline
&GRB     & $T_{90}($s$)$ & $z$ &Ref.$^a$ \\
\hline
&050126  & 27   &  1.29     & 1  \\
&050223  & 23   &  0.5915   & 2--3  \\
&050416A & 2.5  &  0.6535   & 4  \\
&050826  & 35   &  0.297    & 5  \\
&051016B & 4.0  &  0.9364   & 6  \\
&060814  & 145  &  0.84     & 7 \\
&060912A & 6.0  &  0.937    & 8  \\
&061126  & 27   &  1.1588   & 9  \\
&061210  & 85   &  0.4095   & 10-11  \\
&061222A & 96   &  2.088    & 12  \\
\hline
\end{tabular}
\end{center}
\begin{minipage}{18cm}
$^a$ References.
(1)~\citealt{berger05b}; (2)~\citealt{berger05c}; (3)~\citealt{pellizza06}; (4)~\citealt{cenko05}; (5)~\citealt{mirabal07}; (6)~\citealt{soderberg05}; (7)~\citealt{thoene06}; (8)~\citealt{levan07}; (9)~\citealt{perley08}; (10)~\citealt{palmer06}; (11)~\citealt{berger07}; (12)~\citealt{perley09}.
\end{minipage}
\label{tab:sample2}
\end{table}


\end{document}